\documentclass[12pt]{article}
\title{\bf Neutrino Propulsion for Interstellar Spacecraft}  
\author{J. A. Morgan \\
The Aerospace Corporation, El Segundo, CA 90009, U. S. A.}
\date{July 3,1997}
\begin{document}
\maketitle
\begin{abstract} An exotic spacecraft propulsion technology is described 
which exploits parity violation in weak interactions.  Anisotropic neutrino 
emission from a polarized assembly of weakly interacting particles 
converts rest mass directly to spacecraft impulse.
\end{abstract}

\section{Introduction}

The photon rocket has long been a familiar example of special 
relativistic kinematics \cite{ref1}.  The speed of light as the ultimate rocket exhaust 
velocity also surfaces in discussions of interstellar travel \cite{ref2}.  But few 
regard the photon rocket as even remotely practical.  This Note outlines a 
method of spacecraft propulsion which resembles the photon rocket 
closely, and which appears, in principle, capable of reduction to practice.
  
Parity nonconservation in the weak interaction can produce thrust by 
anisotropic neutrino emission from a polarized assembly of weakly 
unstable particles or nuclei.   A conceptual propulsion system, in which 
decay or nuclear capture of muons from the annihilation of bulk antimatter 
generates thrust, illustrates the principles involved.  
 
This scheme may be considered a variant on propulsion methods 
based upon reaction from radioactive decay products, which have appeared 
in the literature at intervals since their proposal by Goddard and 
Tsiolkovsky early in this century \cite{ref3,ref4}.  It is distinct from these in that the decay 
particles are collisionless and electrically neutral, that the anisotropy of 
parity-violating weak decays resolves the otherwise intractable problem of 
collimation, and that the small cross section for neutrino interactions with 
other matter permits thrust generation to take place as a volume, instead of 
as a surface, effect.
\section{Physical Principles}
\subsection{Parity nonconservation and thrust generation}

Consider a nucleus of spin $J$ polarized along $z$ which captures a light 
particle, accompanied by the emission of a neutrino with energy $Q$, 
momentum $q=Q/c$.  The capture imparts to the daughter nucleus 
momentum $\mathbf{q}$ directed opposite to the neutrino, along with a small recoil 
energy.  As the only preferred axis is the parent nuclear spin $\mathbf{J}$, the rate 
may be written as a Legendre polynomial expansion in $\cos(\theta)=\mathbf{q\cdot J}/qJ$:
\begin{equation}
I(\theta)=\sum_{l}a_{l}P_{l}(\cos(\theta)) \label{eqn1}
\end{equation}
								
Because $J$ is an axial vector, while $q$ is a polar vector, their inner 
product is a psuedoscalar, which is odd under parity.  In spherical 
coordinates, the effect of the parity operator is to transform $\theta$ to $\pi-\theta$. 
Terms which do not conserve parity thus appear as odd harmonics in eqn. 
\ref{eqn1}.  As a result, the average impulse per capture along $z$
\begin{equation}
\frac{\int_{4\pi}d\,\Omega\:q_{z}I(\theta)}{\int_{4\pi}d\,\Omega\:I(\theta)}
=\frac{1}{3}\frac{Q}{c}\frac{a_{1}}{a_{0}}\equiv\eta\frac{Q}{c}
\label{eqn2}
\end{equation}
does not vanish.  Only a nonvanishing $a_{1}$ coefficient allows neutrinos to 
transport a net momentum flux.  One concludes that an interaction which 
conserves parity cannot transport a net flux of momentum from a polarized 
assembly of unstable particles, that a decay whose angular intensity contains 
a linear term in $\cos(\theta)$ will violate parity, and that only such an intensity 
pattern can produce thrust.
	
\subsection{Microscopic basis of thrust generation}

The weak processes considered here involve muons produced by 
charged pion decay following proton-antiproton annihilation.  Muons decay 
by emission of electron and muon neutrinos in a three-body process. The 
spectrum of free positive muon decay at rest \cite{ref5} gives $<q_{z}>=Q/10$ for the 
average momentum carried away by neutrinos, or $\eta_{+}=1/10$.  Negative 
muons also undergo nuclear capture, subsequently emitting neutrons in 
over 95 \% of captures.  Collimated neutrino emission results from muon 
capture in which either the muon or a nonzero spin target nucleus is 
polarized. 
 
The conversion efficiency $\eta_{-}$ for negative muon capture by a 
polarized nucleus is calculated following \cite{ref6}, incorporating polarization 
effects after \cite{ref7}.  It will be assumed that capture is dominated by $L=1$ and 
$L=2$ multipoles, that recoil corrections $O(v/c)$ can be neglected and, initially, that 
the nucleus is 100 \% polarized and the muon unpolarized.  
Nuclear matrix elements for individual transitions are computed with a 
modified closure approximation \cite{ref8} in the pure shell model \cite{ref9}.  (The net 
effect of interference terms is estimated to be $\mbox{O(recoil)}$ for medium-mass 
nuclei.)  

In this manner one may estimate  $\eta_{-}$  for transitions to a specified 
final angular momentum and parity.  Its magnitude typically lies between 
0.25 and 0.3 at the upper end, but can take on values near zero.

Detailed calculations of muon capture \cite{ref10} find the capture strength 
concentrated in a small number of final spin/parity states.  $L=1$ transitions 
to states of high spin appear favored, with 65-75 \% of the capture going to 
these states.  It appears reasonable to assume target nuclei exist with net 
efficiency $\eta_{-}$ in the range 0.15-0.25, within recoil corrections $\mbox{O(0.04)}$.  

Hyperfine coupling between the target nucleus and an unpolarized $1S$ 
muon will polarize the muon spin slightly at the expense of the target 
nucleus spin \cite{ref11}.  For high nuclear spins, the polarization loss is small.  
Muon polarization alters the relative importance of transverse and 
longitudinal multipoles \cite{ref12} so as to increase $\eta_{-}$ somewhat, by perhaps 25-
33 \% of its magnitude for unpolarized muons in favorable cases.
  
An estimate for $\eta_{-}$ of ca. 0.2 is taken in the following.

\section{Propulsion System Concept}

The propulsion system operates, in outline, as follows:   
annihilation yields, after free decay of prompt annihilation products, about 
3.16 $\pi^{\pm}$ and 1.85 $\pi^{0}$'s \cite{ref13}.  It is assumed that neutral annihilation products 
are discarded so as to contribute negligibly to waste heat production.  
Muons from pion decay are decelerated to low energies.  Positive muons 
are stopped and (rapidly) polarized, then decay.  A polarized nuclear target 
stops and captures the negative muons, resulting in a neutron flux.  The 
neutrons are moderated and captured in an isotopic mix of nuclei which 
regenerates the initial target isotope.  Both the absorbers which decelerate 
the muons and the thrust-generating interaction regions are coupled to a 
high thermal power, low specific mass waste heat rejection system.

\subsection{Conversion efficiency}

A simple estimate of pion energetics following annihilation results 
from the pion multiplicities on the assumption that pions share the available 
mass-energy equally.  The resulting mean pion kinetic energy is 236 MeV.  
Pions are assumed to decay in flight.  Muon neutrinos from charged pion 
decay carry off about 90 MeV on average.  It is assumed that charge 
separation of muons may be achieved with gradient drift, and that they are 
brought to low kinetic energy by suitable absorbers.  Because of the spread 
in annihilation product kinetic energy, this process requires a degree of 
momentum sorting.

The fraction of proton pair mass that appears as muon rest mass is
\begin{equation}
\alpha=\frac{n_{\pi_{\pm}}m_{\mu}}{2m_{p}}  \label{eqn3}
\end{equation}								
(in an obvious notation), divided equally between charge states.  Combining 
the conversion efficiency for free positive muon decay with that for 
negative muon capture, adjusted for losses from residual nuclear excitation, 
leads to an overall, ideal efficiency estimate
\begin{equation}
\eta=\frac{\alpha}{2}[\eta_{+}+0.85\cdot\eta_{-}]\cong0.025.
\label{eqn4}
\end{equation}	
\subsection{Muon interactions}

Stopped positive muons form muonium with unpaired electrons in 
most substances.  It is assumed that dynamical polarization by muonium 
hyperfine coupling in a strong magnetic field polarizes muons rapidly 
compared to their lifetime, whereupon their free decay imparts impulse. 
The thermal energy deposited in the ultimate target medium by a stopping 
muon should not, therefore, exceed the hyperfine splitting.

Negative muons are made to stop in a highly polarized target of a 
suitable nucleus such as $^{51}V$ or $^{55}Mn$, whose most probable muon capture 
daughters are all stable.  In most targets negative muons rapidly decelerate 
to rest, whereupon they form muonic atoms.  In medium-mass nuclei 
($Z>20$ or so), they overwhelmingly undergo nuclear capture from the 
muonic $1S$ state instead of free decay \cite{ref14}.

The target may be gaseous \cite{ref15}.  Negative muons can deposit 
significant heat in the capture target without compromising nuclear 
polarization, and thus only require coarse momentum sorting.  In fact, 
muon capture probably must occur in a confined plasma, because hyperfine 
coupling between $K$ and $L$-shell atomic electrons in the target and the $1S$ 
muon-also subject to hyperfine coupling with the target nucleus-poses a 
risk of catastrophic loss of target nuclear polarization.  A thermal energy 
of 0.5 keV, or just less than $5 \times 10^{6}$ K, suffices for most medium-mass 
nuclei \cite{ref11}.
	
\subsection{Neutron issues}

The daughter nucleus deexcites mainly by neutron emission.  It is 
proposed to recover the target nucleus for subsequent reuse by operating a 
catalytic neutron capture chain on daughter nuclei.

Recovery begins with moderation of the neutrons to thermal energies 
in a low-neutron absorption moderator such as graphite or $D_{2}\,^{16}O$.  A jacket 
composed of lighter isotopes of the daughter nucleus captures the neutrons, 
thus running a neutron capture chain terminating in an isobar of the 
original nucleus that quickly decays back to the parent isotope.  Chemical 
separation of the parent isotope completes the chain.

Proton and alpha channels cause some leakage out of the chain.  
However, the nuclei are not lost to the system.  Daughter nuclei from $(\mu_{-}
,np)$ or $(\mu_{-},\alpha)$ reactions lie on the neutron-rich side of stability and will 
consequently $\beta$-decay back to the stability line.  An extra neutron capture 
or two reinserts them into the main chain.  Neutrons to supply this loss 
come from diverting a fraction of the negative muon flux to capture in 
heavy nuclides whose composition is otherwise immaterial.  An estimate 
based upon the neutron multiplicity per muon capture \cite{ref14} indicates that a 
0.01 penalty to $\eta$, of recoil order, suffices to redress a 3 \% nucleon deficit 
per capture.

\subsection{Waste Heat Rejection}

For purposes of discussion it will be assumed that waste heat is 
rejected by liquid droplet radiators.  Nordley~\cite{ref16} estimates that a specific 
mass of $10^{-5}\: kg/W_{t}$ at an outlet temperature of 900 K represents an 
attainable extension of liquid droplet radiator performance.  In the present 
instance, in which no penalty accrues to high operating temperatures, it 
seems permissible to conjecture that radiators may use as a working fluid a 
refractory metal at an outlet temperature in the low 2000 K's.  If 
evaporation losses are low enough, and the nonradiating mass depends 
weakly upon temperatures, then the specific mass of a liquid drop radiator 
scales as a function of the working fluid inlet and outlet temperatures only 
\cite{ref17}.  Estimates for a variety of candidate working fluids indicate specific 
masses of order $10^{-7}\: kg/W_{t}$ can probably be attained, but extension to 
greatly lower values appears difficult.

\subsection{Spacecraft dynamics}

Consider a simple rocket propelled by neutrinos.  A change in 
spacecraft rest mass $dM$ yields an impulse $\eta c\,dM$ and discarded energy $(1-\eta)c^{2}\,dM$.  
Let $Nt$ annihilations in a time $t$ impart impulse to the spacecraft at a rate
\begin{equation}
\dot{p}=2Nm_{p}c\eta, \label{eqn5}
\end{equation}	
accompanied by local waste heat power
\begin{equation}
\dot{W} \equiv N\delta W \label{eqn6}
\end{equation}
within it, where
\begin{equation}
  \delta W\cong \left[2m_{p}c^{2}\!-n_{\pi^{\pm}}\!
\left[m_{\mu}c^{2}+\gamma_{0}(m_{\pi^{\pm}}-m_{\mu})c^{2}\right]
	\!-\!\gamma_{0}n_{\pi^{0}}m_{\pi^{0}}c^{2}\!+\!E_{0}\right].  
\end{equation}						
The average Lorentz factor $\gamma_{0}$  is 2.7.  $E_{0}$ includes ca. 38 MeV 
from residual nuclear excitation, recoil energy from $\mu^{+}$ decay, and positron 
annihilation.  Waste heat production per annihilation is about 690 MeV.  
The propulsion system mass is
\begin{equation}
M=\dot{W}\sigma_{W}+m_{mod}+m_{abs}+m_{plant}, \label{eqn7}
\end{equation}	
where  $\sigma_{W}$ is the specific mass for waste heat rejection, and the masses of 
neutron moderator and catalytic chain absorber (governed by neutron 
scattering and absorption lengths, respectively) are, along with those of other propulsion plant 
components, assumed to scale weakly with power.  The high waste heat 
power levels expected in this concept suggest examining the extreme in 
which the propulsion system fixed mass is dominated by the waste heat 
rejection system.  In this limit, the final acceleration is given by the ratio of 
eqns. (\ref{eqn5}) and (\ref{eqn7}):
\begin{equation}
a=2\frac{\eta m_{p}c}{\delta W \sigma_{W}}\equiv\epsilon\eta.
\label{eqn8}
\end{equation}	
With this definition of $\epsilon$ the spacecraft expends mass at the constant rate
\begin{equation}
\frac{\dot{M}}{M_{0}}=\frac{\epsilon\mu_{f}}{c}. \label{eqn9}
\end{equation}  Its rapidity $\rho\equiv\tanh^{-1}(v/c)$ obeys the traditional rocket equation [1,2],
\begin{equation}
\rho=-\eta\,\log(1-\frac{\epsilon\mu_{f}}{c}t).
\label{eqn10}
\end{equation}  Here $1-\epsilon\mu_{f}t/cº\equiv\mu(t)$ is the inverse mass ratio; its final value is 
$M(t_{f})/M_{0} =\mu_{f}$.  For $\sigma_{W}=10^{-7}\:kg/W_{t}$,$\,\epsilon=9.1 cm/s^{2}$.
  The distance traveled in proper time t is
  \begin{equation}
  s=c\int_{1}^{\mu_{f}}\beta(\mu(t))\frac{dt}{d\mu}\:d\,\mu
  =-\frac{c^{2}}{\epsilon\mu_{f}}\int_{1}^{\mu_{f}}\beta(\mu)\:d\,\mu;
  \label{eqn11}
  \end{equation}
  \begin{equation}
  s(\mu_{f})=\frac{\eta c^{2}}{\epsilon}\frac{1}{\mu_{f}}\left[\mu_{f}\log(\mu_{f})-\mu_{f}+1\right],
  	\label{eqn12}
  \end{equation}	
substituting $\rho$ for $v/c\equiv\beta,$ since we are concerned with small rapidities 
$\leq0.1$. 
 
In addition to the classical rocket, two other concepts are examined:
	
\subsubsection{Ram augmentation}  In this variant the spacecraft carries its 
antiprotonic fuel, but obtains protons from the interstellar medium \cite{ref18}.  
Four-momentum balance in the rocket rest frame at $t=0$ gives
\begin{equation}
-\frac{d\,M}{M}\cong\frac{d\,\rho}{\exp(-\rho)+2\eta-1},
\label{eqn13}
\end{equation}
which integrates to
\begin{equation}
\log(\mu)=\frac{1}{1-2\eta}\left[\rho+\log\left[\frac{\exp(-\rho)+1-2\eta}{2\eta}\right]\right],
\label{eqn14}
\end{equation}
a transcendental equation for $\rho(\mu)$.  The expression for ram deceleration is 
obtained by changing the sign of $\rho$ in eqn. (\ref{eqn13}).

\subsubsection{Pellet stream}The limit of ram augmentation occurs when a stream
of fuel pellets is launched in advance of the spacecraft with a tailored 
profile of pellet speeds and launch times, so that the spacecraft ingests each 
fuel pellet at zero velocity in its proper frame \cite{ref18}.  For the pellet stream,
\begin{equation}
\log(\mu)=\frac{1}{2\sqrt{\eta-\eta^{2}}}\left[\arcsin\left[
\frac{1+(2\eta-1)\cosh(\rho)}{\cosh(\rho)+(2\eta-1)}\right]-\frac{\pi}{2}\right].
\label{eqn15}
\end{equation}

Modifications to eqn. (\ref{eqn6}) for ram-augmented systems are second 
order in $v/c$.  For the augmented variants, the distance traveled is computed 
by numerical integration of eqn. (\ref{eqn11}) but with $\epsilon$ replaced by $\epsilon/2$
  in order 
to maintain consistency with the simple rocket.

\section{Performance Estimates}
	
Consider a rendezvous mission with a star at a distance of 10 pc, and 
with a duration held fixed at 1000 years.  Table 1 displays mission profiles 
for hybrids of pure neutrino rocket, ram- and pellet-augmented rocket, and 
photon sail acceleration at 0.3 gravities \cite{ref18}.

There is a critical value of $\epsilon$ below which it is impossible to travel 10 pc in 
1000 years.  The ram/ram profile requires $\epsilon_{c}=23.6$.  Other profiles in the table 
have smaller critical values.  The photon beam/ram deceleration profile evidently 
has the greatest latitude in this regard; for $\eta=0.025$, $\epsilon_{c}=2.0$.  If 
the efficiency is as 
low as $\eta=0.015$, $\epsilon_{c}=3.2$.  The mass ratio for this last example is 2.64.

\section{Discussion}

The propulsion system concept outlined above delivers 
semirelativistic rapidity changes for modest mass ratios, but does so at 
comparatively low efficiency.  The inefficiency arises from the small 
charged pion multiplicity from   annihilation at rest.  Purely neutrino-
powered examples have been described to illustrate the principle most 
directly, and for comparison with the classical photon rocket.  One would 
doubtless prefer to treat that portion of annihilation energy which cannot 
be converted into collimated muon neutrinos as something other than waste 
heat; to generate thrust with it!  The neutrino rocket may furnish a topping 
cycle for some other, presumably more efficient, method of annihilation-based propulsion.

\begin{table}
Table 1.  Sample Rendezvous Profiles 
{\small \begin{tabbing}
Method\=.................\=photon/ram    \=photon/rocket   \=pellet/ram   \=pellet/rocket   \=ram/ram \kill
Method\>\>photon/ram\>photon/rocket\>pellet/ram\>pellet/rocket\>ram/ram \\
acc. mass ratio\>\>N/A\>N/A\>2.35\>1.84\>5.13 \\
acc. t(yr)\>\>0.11\>0.10\>290\>179\>339   \\
acc. distance(pc)\>\>$6.1 \times 10^{-4}$\>$4.4 \times 10^{-4}$\>1.63\>0.752\>2.10  \\
max. v/c\>\>0.035\>0.030\>0.043\>0.030\>0.041 \\
coast distance(pc)\>\>9.06\>7.52\>6.91\>5.30\>7.39  \\
coast t(yr)\>\>849\>755\>525\>557\>593   \\
dec. mass ratio\>\>1.70\>3.28\>1.86\>3.44\>1.82  \\
dec. t(yr)\>\>151\>245\>185\>263\>68    \\
dec. distance(pc)\>\>0.94\>3.13\>1.46\>3.95\>0.50  \\
total mass ratio\>\>1.70\>3.36\>4.37\>6.32\>9.36  \\
\end{tabbing}} 
 $\eta_{0} = 0.025$.  $\epsilon=9.0$, except for ram/ram ($\epsilon=23.6$). Photon sail profiles use an acceleration 
of 0.3 gravities.
\end{table}
\end{document}